\documentclass[aps,preprint,amsmath,amssymb]{revtex4}
\usepackage{graphicx}
\begin{document}
\begin{center}
{\large\bf 
Quantum Criticality and 
the Kondo Lattice
}
\\[0.5cm]

Qimiao Si
\\

{\em Department of Physics and Astronomy, Rice University,
Houston, TX 77005, USA}\\


\end{center}

\vspace{0.5cm}
{Quantum phase transitions (QPTs) arise as a result of competing interactions
in a quantum many-body system. Kondo lattice models, containing a
lattice of localized magnetic moments and a band of conduction electrons,
naturally feature such competing interactions. A Ruderman-Kittel-Kasuya-Yosida
(RKKY) exchange interaction among the local moments promotes magnetic
ordering.
However, a Kondo exchange interaction between
the local moments and conduction electrons favors the Kondo-screened
singlet ground state.

This chapter summarizes the basic physics of QPTs
in antiferromagnetic Kondo lattice systems.  Two types of quantum critical
points (QCPs) are considered. Spin-density-wave quantum criticality
occurs at a conventional type of QCP, which invokes only the fluctuations
of the antiferromagnetic order parameter. Local quantum criticality describes
a new type of QCP, which goes beyond the Landau paradigm
and involves a breakdown of the Kondo effect. 
This critical Kondo breakdown effect leads to non-Fermi
liquid electronic excitations, which are part of the critical 
excitation spectrum and are in addition to the fluctuations 
of the magnetic order parameter. Across such a QCP, there is a sudden
collapse of the Fermi surface from large to small.
I close with a brief summary of relevant experiments, and outline a number 
of outstanding issues, including the global phase diagram.
}



\newpage

\section{Introduction}

\subsection{Quantum Criticality: Competing Interactions in
Many-body Systems}

Quantum criticality describes the collective fluctuations associated
with a second order phase transition at zero temperature.  It occurs
in many-body systems as a result of competing interactions that
foster different ground states. An example is the transverse-field
Ising model:
\begin{eqnarray}
H = - I \sum_{<ij>}\sigma_i^z \sigma_j^z -h_t \sum_i \sigma_i^x
-h \sum_i \sigma_i^z .
\label{QS:eqn:tfim}
\end{eqnarray}
Here $\sigma_i^z$ and $\sigma_i^x$ denote the $z$- and 
$x$-components of a localized spin-$1/2$ magnetic moment at
site $i$, $I$ is the
Ising exchange interaction
between the $z$-components of nearest-neighbor spins,
and $h_t$
is an external magnetic field
applied along the transverse 
$x$-direction with 
$g\mu_B$ set to $1$.
In order to probe the
spontaneous symmetry breaking, a longitudinal field $h$ is also
introduced which will be set to $0^+$ at the end of calculation.
In one spatial dimension, this model can be exactly
solved through a Jordan-Wigner transformation~\cite{QS:Pfeuty}.

\begin{figure}[t]
\centerline{\includegraphics[width=0.7\linewidth,height=0.35\linewidth]{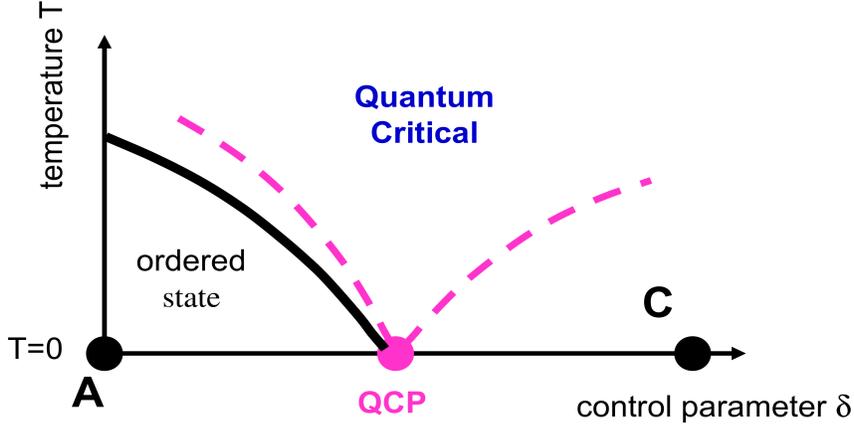}}
\caption{
Phase diagram of the transverse-field Ising model.
$\delta$ is a non-thermal control parameter, which tunes 
quantum fluctuations.
The solid line denotes phase transitions into 
the ordered state, and the corresponding 
transition temperature approaches zero as $\delta$ is tuned towards
the QCP. At non-zero temperatures, a quantum-critical regime arises;
the dashed lines describe the crossovers into and out of 
this regime.
}
\label{QS:tfim}
\end{figure}

The $I$ and $h_t$ terms represent competing interactions of the
system. Their competition can be parametrized in terms of a
dimensionless quantity, $\delta \equiv h_t/I$.
We consider the model in the $T-\delta$ phase diagram, Fig.~\ref{QS:tfim},
starting from
the readily solvable points $A$ and $C$.
Point $A$ corresponds to $T=0$ and a vanishing transverse field, $\delta=0$,
where we want to minimize the exchange energy.
The ground state is 
$\Pi_i|\!\!\uparrow\rangle_i$, 
with all the spins lined up
along the positive $z$ direction. It spontaneously breaks
a global $Z_2$ symmetry: under the operation
$\sigma_i^z \to -\sigma_i^z$, for every
site $i$, the ground state is changed even though the Hamiltonian is
invariant for $h=0$. The macroscopic order is described by Landau's
order parameter,
defined as
\begin{eqnarray}
\phi \equiv
\lim_{h\to 0^+}
\lim_{N_{\mathrm{site}}\to \infty} M/N_{\mathrm{site}} ,
\label{QS:order-parameter-def}
\end{eqnarray}
where $M=\langle \sum_i\sigma_i^z\rangle$ and $N_{\mathrm{site}}=\sum_i$. We reach
an important conclusion that
\begin{eqnarray}
\phi
= 1 , ~~~ {\rm at~point~A}.
\label{QS:tfim-order-parameter-pointA}
\end{eqnarray}
Point C is also at $T=0$ but has $\delta \gg 1$. Since the
transverse field is the largest coupling ($h_t \gg I$),
the system minimizes the internal energy by lining up all
the spins along the positive $x$ direction. The
ground state is $\Pi_i|\!\rightarrow\rangle_i$.
At each site,
$|\!\rightarrow\rangle = (|\!\uparrow\rangle + |\!\downarrow\rangle)
/\sqrt{2}$ maximizes the tunneling between the spin up and
spin down states.
These two spin states have an equal probability,
and the order parameter vanishes:
\begin{eqnarray}
\phi =0, ~~~{\rm at~point~C}.
\label{QS:tfim-order-parameter-point-C}
\end{eqnarray}
When we tune the non-thermal control parameter $\delta$ from
point $A$ to point C, we can expect at least one phase transition
separating the 
magnetically
ordered 
($\phi \neq 0$) and
disordered ($\phi =0$) states.

In the absence of the transverse field, the Hamiltonian involves only
$\sigma_i^z$, which all commute with each other.  
In the presence of the transverse field,
$\sigma_i^x$ also appears in the Hamiltonian, which now 
contains variables that do not 
commute with each other;
the system
becomes quantum-mechanical. Varying $\delta$ amounts to tuning
the degree of quantum tunneling or, equivalently, the degree to
which the zero-point motion is manifested in the many-body properties.
This is referred to as tuning the degree of 
\textit{quantum fluctuations}.

Fig.~\ref{QS:tfim} illustrates the $T-\delta$
phase diagram of the model in dimensions
higher than one.
In one dimension, the line of the finite-temperature phase transitions
collapses to the zero-temperature line, in accordance with the
Mermin-Wagner theorem.
Importantly, the zero-temperature transition is continuous,
with the 
$T=0$
order parameter $\phi$ smoothly going to zero as the
control parameter $\delta$ is increased to $\delta_c$, the
QCP. 
In common with
its classical counterpart
at non-zero temperatures, a QCP features critical fluctuations.
These collective fluctuations are responsible for
non-analyticities in the free energy as a function of $\delta$ and
other parameters.

Historically, the 
first example in which such QCPs
were formulated in the modern language of critical
phenomenon is the case of metallic paramagnets
undergoing a second-order phase transition at zero temperature
into a Stoner ferromagnet or spin-density-wave (SDW)
antiferromagnet. 
Microscopically,
Hertz~\cite{QS:Hertz} 
modeled this transition with 
a one-band Hubbard model.
The Coulomb repulsion among the electrons favors magnetic
ordering, and the electrons' kinetic energy induces paramagnetism.
The result is an itinerant magnetic QCP
or,
in the case of antiferromagnetic (AF) order, 
an
SDW QCP.
Hertz constructed an effective field theory for
the fluctuations of the order
parameter, $\phi$, in both space ($\bf x$) and imaginary
time ($\tau$). 
The result is a quantum Ginzburg-Landau action,
\begin{eqnarray}
{\cal S}
\!  = \!\!
\int d {\bf q}~
\! \!
\frac{1}{\beta}
\sum_{i\omega_n}
(r + c {\bf q}^2 + |\omega_n|/\Gamma_{\bf q})
\, {\bf \phi} ^2
+
\int \!\!~
\!\!\, u \, {\bf \phi} ^4
+ \ldots .
\label{QS:S-Hertz}
\end{eqnarray}
The effective dimension of the fluctuations is $d+z$, where $d$ is
the spatial dimension and $z$, the dynamic exponent, counts the
effective number of extra spatial 
dimensions 
that the temporal
fluctuations correspond to.
In Eq.~(\ref{QS:S-Hertz}),
the quadratic part
is expressed 
in terms of wavevector
${\bf q}$ and Matsubara frequency $\omega_n$, 
which are reciprocal to ${\bf x}$ and $\tau$, respectively.
At non-zero temperatures, this action
gives rise to a quantum-critical 
regime\cite{QS:Sachdev-book,QS:Chakravarty,QS:Millis,QS:Moriya},
whose physical properties are controlled by the many-body 
excitations of the system's ground state 
at the QCP.

What underlies the Hertz description is the Landau notion
that fluctuations of the order parameter 
are
the only critical
degrees of freedom.
Order parameters, as we encounter here, are classical variables.
However, 
theoretical developments on 
QCPs
of
heavy fermion metals~\cite{QS:Si-Nature,QS:Colemanetal}
and insulating quantum magnets~\cite{QS:Senthil-dcqp} have
shown that this is not the only possibility. Instead,
inherent quantum modes can emerge as part of the critical degrees
of freedom. Identifying these additional modes is 
nontrivial,
as the Landau paradigm of doing so using symmetry-breaking patterns
can no longer be used. This task must be completed before the
critical field theory can be constructed.

\subsection{Heavy Fermion Metals}

Heavy fermion metals refer to 
rare-earth- or actinide-based
intermetallic compounds in which the
effective mass of the electronic excitations is
hundreds 
of times 
the bare electron mass.
They typically 
arise in compounds including Ce, Yb, and U that 
contain partially-filled 4$f$ or 5$f$ 
orbitals.
The large effective mass 
originates
from strong electron correlation,
i.e.,
a large ratio of the on-site Coulomb repulsive
interaction to the kinetic energy.
Among the 
outer-shell 
orbitals of a rare-earth or actinide ion,
the $f$-orbitals are closer to the origin.
Correspondingly, the average distance
among the $f$-electrons occupying 
the 
same site is relatively short,
leading to the enhanced Coulomb interaction.

Microscopically, we have a narrow $f$-electron band coupled
to some wider conduction-electron bands through a finite
hybridization matrix~\cite{QS:Hewson}. In many compounds, the $f$-electrons
are so strongly correlated that their valence 
occupation 
stays at
an integer value. This would be one 
4$f$-electron
for Ce-based compounds, and thirteen 
4$f$-electrons,
or,
equivalently, one 
4$f$-hole, 
for Yb-based compounds.
This integer-valency can be thought
of in terms of the $f$-electron band being in its Mott-insulating
state, representing an example of the 
\textit{orbitally-selective
Mott insulator}.
At energies much smaller than the gap of this orbitally-selective
Mott insulator, the $f$-electrons no longer possess
charge fluctuations and behave as localized magnetic moments.

\begin{figure}[t]
\centerline{\includegraphics*[width=0.7\linewidth,height=0.24\linewidth]{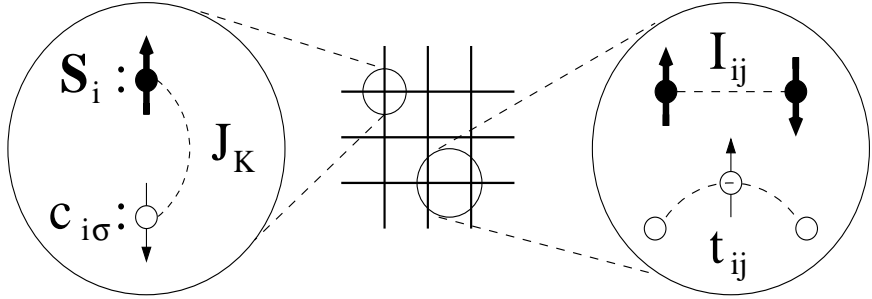}}
\caption{
The Kondo lattice model.
At each site of the lattice,
there are two types of microscopic degrees of freedom:
a spin-$1/2$ local moment ${\bf S}_i$ and a conduction electron 
$c_{i\sigma}$, 
which are coupled to each other through 
an antiferromagnetic Kondo exchange interaction,
$J_K$.
The local moments 
interact with each other through 
a pair-wise exchange interaction $I_{ij}$. 
The conduction electrons have a tight-binding 
hopping matrix $t_{ij}$. 
}
    \label{QS:kondo-lattice}
\end{figure}

What results is a Kondo lattice Hamiltonian, illustrated in
Fig.~\ref{QS:kondo-lattice}. It contains a lattice of local
moments and,
for simplicity, 
one
band of conduction electrons.
The local moments interact with each other through the RKKY
exchange coupling,
and they interact with the conduction electrons via an
AF Kondo exchange coupling. The fact that these interactions
are much smaller than the orbitally-selective Mott insulating gap
is an important feature of the Kondo lattice model.
This separation of energy scales allows the
local moments to be easily observable, as it gives rise to an
extended temperature window over which the local moments are
manifested in a Curie-Weiss form for the bulk spin
susceptibility.

The central microscopic question is how these local
moments interplay with the conduction-electron bands.
The early 1980s saw extensive studies of this problem,
which were built upon the historical work on the
single-impurity Kondo effect. These studies showed
how 
Kondo singlets are formed in the ground state,
and how such Kondo screening gives rise
to 
Kondo resonances in the excitation spectrum. The
resonances are spin-1/2 and charge-e
excitations, and they combine with 
conduction electrons 
to form
the heavy 
quasi-particles. 
This is the picture of 
the 
\textit{heavy Fermi liquid}
~\cite{QS:Hewson}.
Because of the large effective mass, Fermi-liquid
effects become amplified.
The sentiment was that
heavy fermion metals represent a prototype system for Landau's
Fermi liquid theory of charged fermions in the presence of a lattice.

That 
RKKY interactions and Kondo interactions compete against each
other was already recognized early on~\cite{QS:Doniach, QS:Varma76}.
However, the Fermi
liquid theory of the paramagnetic heavy-fermion systems was considered to
be so successful that the magnetic order was mostly considered as
a descendant of a heavy Fermi liquid. In this picture,
an antiferromagnetically ordered heavy fermion metal arises from
an RKKY-interaction-induced SDW instability of the heavy
quasi-particles 
near their Fermi surface.
Likewise, heavy-fermion superconductors are considered in terms
of the Cooper-pairing instability of the same
quasi-particles.

\subsection{Quantum Critical Point in Antiferromagnetic Heavy Fermions}

The field of heavy fermions was interrupted by the
discovery and extensive studies of high temperature cuprate
superconductivity. When the field re-emerged
in full force, the focus was changed
in an important way. It was recognized that heavy fermion
metals represent an important testing ground for the
breakdown of Fermi liquid theory in general~\cite{QS:Maple.94},
and the nature of QCPs in particular.

Over the past decade, magnetic heavy fermion metals have become
a prototype setting to realize and explore QCPs.
The list of quantum critical heavy fermion metals is by now
relatively long, and the readers are referred to
some more 
comprehensive
reviews~\cite{QS:Gegenwart.08,QS:HvL-RMP}
and Chap.~18\cite{chapter:gegenwart}
for a more extensive discussion on the materials and experimental
aspects.
The most prominent examples are
CeCu$_{\rm 6-x}$Au$_{\rm x}$, YbRh$_{\rm 2}$Si$_{\rm 2}$,
CePd$_{\rm 2}$Si$_{\rm 2}$,
and
CeRhIn$_{\rm 5}$.
Quantum criticality is being explored
in many
condensed matter systems.
Arguably, heavy fermion
QCPs
have been the most systematically studied, 
because such QCPs have been 
explicitly identified
in a number of available materials.

One of the first hints about the failure of the
order-parameter-fluctuation picture for quantum criticality
came from 
inelastic neutron-scattering measurements~\cite{QS:Schroder}
in CeCu$_{\rm 6-x}$Au$_{\rm x}$. The inelastic neutron
scattering cross section measures the dynamical spin susceptibility,
$\chi({\bf q},\omega,T)$. For ${\bf q}={\bf Q}$,
$\chi({\bf Q},\omega,T)\propto 1/(-i\omega)^{\alpha}$
at
$\omega \gg T$, with a non-mean-field exponent
$\alpha \approx 0.75$. The static susceptibility goes as
$\chi({\bf q},\omega=0)
\propto 
1/(\Theta_{\bf q} + a T^{\alpha})$,
with the same exponent $\alpha$ as seen in
$\chi''({\bf Q},\omega,T)$, and with a Weiss temperature
$\Theta_{\bf q}$ which goes to zero as ${\bf q}$ approaches
${\bf Q}$. Finally, $\chi({\bf Q},\omega)$ satisfies
$\omega/T$ scaling.
In contrast, 
the SDW QCP has a dynamic exponent $z=2$,
and an effective dimensionality of the order-parameter
fluctuations $d+z \ge 4$ for d=2,3. The critical theory,
Eq.~(\ref{QS:S-Hertz}),
describes a Gaussian
fixed point. The critical exponents are expected
to take the mean-field value $\alpha=1$. Moreover,
the non-linear interactions that give
rise to spin damping must vanish. In other words, the effective
interaction $u(T)\sim T^{\theta}$,
with $\theta > 0$,
and hence the damping
rate 
scale
as $\Gamma(T) \sim T^{1+\theta}$. Frequency 
appears 
with temperature in the combination
$\omega/\Gamma(T)$, resulting in a violation of
$\omega/T$ scaling.\footnote{
This 
persists even when 
the effect
of non-analytic corrections, which arise in the process
of integrating out the fermions~\cite{QS:Abanov_Chubukov04,QS:belitz-rmp05},
is taken into account.}

To search for new kinds of QCPs, 
the question is what types of new
quantum critical modes emerge. For heavy fermion metals,
these modes are characterized by a critical Kondo
breakdown~\cite{QS:Si-Nature, QS:Colemanetal}.
The notion is that, at the
boundary of the AF order, the amplitude of the
Kondo 
entanglement\footnote{That is, the strength of the spin singlet
formed between the local moments and conduction
electrons; see also Eq.(\ref{QS:Kondo-singlet}) below.}
is severely reduced or
even completely suppressed. A critical suppression of this singlet
amplitude yields new 
types 
of critical modes, which drastically
modify the critical behavior of the spin dynamics,
among other physical properties.

The remainder of the chapter is organized as follows. In Section
\ref{QS:sec:heavy_fl}, we describe the paramagnetic heavy 
Fermi
liquid
state of the Kondo lattice. Our emphasis is the Kondo singlet
formation, and the ensuing development of a 
large
Fermi surface. Section \ref{QS:sec:qcp} is devoted to the nature
of 
QCPs, 
with an emphasis on 
the Kondo breakdown as it appears in
the local quantum criticality.
In Section \ref{QS:sec:af_kl}, we will study
the antiferromagnetically ordered 
part of the phase diagram,
showing that antiferromagnetism can destroy the Kondo effect
and yield a 
small Fermi surface.
Section \ref{QS:sec:global}
considers
the global phase
diagram. 
In Section \ref{QS:sec:expt-AF},
we briefly summarize the relevant experiments.
Some directions for future work are outlined in
Section \ref{QS:sec:sum}.
The second half of the chapter in part overlaps with
Ref.~\cite{QS:Si_PSS.10},
to which I refer for a more complete set of references.

\section{Heavy Fermi Liquid of Kondo Lattices}
\label{QS:sec:heavy_fl}

\subsection{Single-impurity Kondo Model}

To introduce the Kondo effect, we follow the historical route
and first consider the single-impurity Kondo model.
It describes a local moment, ${\bf S}$, interacting
with a band of conduction electrons, $c_{{\bf k}\sigma}$:
\begin{eqnarray}\label{QS:eq:Kondo_impurity}
H_{\rm Kondo} = \sum_{{\bf k}} \varepsilon^{\phantom\dagger}_{\bf k}
c^{\dagger}_{{\bf k}\sigma} c^{\phantom\dagger}_{{\bf k}\sigma}
~+~
J_K {\bf S} \cdot {\bf s}_{c,0} .
\end{eqnarray}
Here,
${\bf s}_{c,0} = (1/2) c_0^{\dagger} \vec{\sigma} c_0$
is the spin
of 
the conduction electrons at the
impurity site, ${\bf x}=0$, 
where
$\vec{\sigma}$ denote the Pauli matrices 
and 
$\varepsilon_{\bf k}$
is the energy dispersion of the conduction electrons.
It is important that the Kondo coupling, $J_K>0$,
is antiferromagnetic.

To address the effect of Kondo coupling, $J_K$, we ask what its
scaling dimension is. This is in the spirit of perturbatively treating
the $J_K$ coupling, and the reference point of our analysis corresponds
to a free spin and a free band of conduction electrons. Recognizing
that the 
autocorrelator 
of a free spin is a constant, we have the
scaling dimension $[{\bf S}]=0$. Also, for free conduction electrons,
the autocorrelator
of $\bf{s}_{0,c}$ 
go 
as $1/\tau^2$, so
$[{\bf s}_{0,c}]=[1/\tau]$. Correspondingly, 
$\int d\tau {\bf S}(\tau)\cdot {s}_{0,c}(\tau)$ 
has a scaling dimension $0$,
and $J_K$ is marginal in the renormalization group (RG) sense.
A well-known 
loop-correction
calculation~\cite{QS:Hewson}
shows that, at the one loop
order, the RG beta function is
\begin{eqnarray}
\beta(J_K) = J_K^2 .
\label{QS:beta-function-Kondo}
\end{eqnarray}
In other words, the AF Kondo coupling is marginally relevant.

The Kondo
coupling renormalizes towards strong coupling as the energy is lowered.
We interpret this as implying that
the effective Kondo coupling is infinite at the fixed point.
This interpretation is verified by a host of studies using
a variety of methods, including the exact solution based on
the Bethe Ansatz and the analysis 
using
conformal invariance.
The infinite coupling at the fixed point describes the physics that
the local moment and the spin of the conduction electrons are locked
into a singlet:
\begin{eqnarray}
|{\rm Kondo~singlet}\rangle=\frac{1}{2}(|\!\uparrow\rangle_f|\!\downarrow\rangle_{c,FS}
-|\!\downarrow\rangle_f|\!\uparrow\rangle_{c,FS} ) ,
\label{QS:Kondo-singlet}
\end{eqnarray}
where $|\sigma\rangle_{c,FS}$ is a linear superposition of the conduction-electron
states near the Fermi energy.

Eq.~(\ref{QS:Kondo-singlet}) is 
an
entangled state between the local moment
and the spins of the conduction electrons. As a result of this entanglement,
the local moment is converted into a Kondo
resonance in the excitation spectrum.
The latter possesses the quantum number of a bare electron,
spin 1/2 and charge $e$. The system is a Fermi liquid, 
in the sense
that low-lying excitations are Landau 
quasi-particles.
The Kondo resonance occurs below a crossover temperature scale,
the Kondo temperature
\begin{eqnarray}
T_K^0 \approx \rho_0^{-1} \exp(-1/\rho_0J_K) .
\label{QS:Kondo-temperature}
\end{eqnarray}

\subsection{Kondo Lattice and Heavy Fermi Liquid}

The microscopic model for heavy fermion materials is the
Kondo lattice Hamiltonian
(Fig.~\ref{QS:kondo-lattice}):
\begin{eqnarray}\label{QS:eq:KLHamiltonian}
H_{\rm KL}=
 \sum_{ ij } t_{ij}
c^{\dagger}_{i\sigma}
c^{\phantom\dagger}_{j\sigma}
+
 \sum_{ ij } I_{ij}
{\bf S}_i \cdot {\bf S}_j
+  \sum_{i} J_K {\bf S}_i \cdot c^{\dagger}_{i}
\frac{\vec{\sigma}}{2} c^{\phantom\dagger}_{i} .
\label{QS:kondo-lattice-model}
\end{eqnarray}
The model contains one conduction-electron band, $c_{i\sigma}$,
with hopping matrix $t_{ij}$,
and,
correspondingly, band dispersion
$\varepsilon^{\phantom\dagger}_{\bf k}$.
At each site $i$, the spin of the conduction
electrons, ${\bf s}_{c,i} = (1/2) c_{i}^{\dagger} \vec{\sigma} c_i$,
is coupled to the spin of the
local moment, $\bf {S}_i$,
via an AF
Kondo exchange interaction $J_K$.

The Kondo effect is one primary mechanism in the Kondo lattice
Hamiltonian to suppress the tendency of the local moments to develop
AF order. In the RG sense, the Kondo coupling still
renormalizes towards strong coupling, leading to the formation of
Kondo singlets. Like in the single-impurity Kondo problem,
this Kondo entanglement in the ground state supports Kondo resonances
in the excitation spectrum. 
However, 
in contrast to the single-impurity case,
the number of the Kondo resonances, being one per site,
is thermodynamically finite, and this will influence the electronic
structure in a
drastic
way.

For concreteness, consider that the conduction
electron band is filled with $x$ electrons per site,
or, equivalently,
per unit cell;
without loss of generality,
we take $0<x<1$. The conduction electron band and the Kondo resonances
will be hybridized, resulting in a count of $1+x$ electrons per site.
The Fermi surface of the conduction electrons alone would therefore
have to expand to a size that encloses these $1+x$ electrons.
This is the large Fermi surface.

\begin{figure}[t]
\centerline{\includegraphics*[width=0.5\linewidth,height=0.20\linewidth]{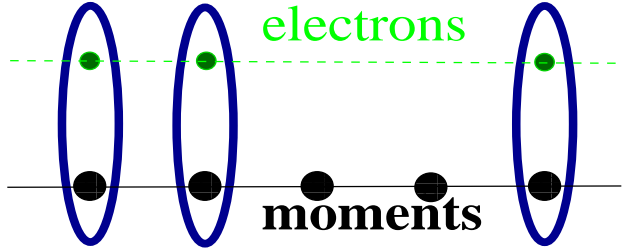}}
\caption{
Kondo singlets in the large Kondo-interaction limit.}
    \label{QS:kondo-singlet}
\end{figure}

To see how this might happen, consider the extreme limit of
$J_K \gg W \gg I$, where $W$ is the width of conduction electron
band, $\varepsilon^{\phantom\dagger}_{\bf k}$,
and $I$ 
is
the typical exchange interaction among the local moments.
This limit is  illustrated in Fig.~\ref{QS:kondo-singlet}.
At each of the $xN_{\rm site}$ sites,
where $N_{\rm site}$
is the number of unit cells in the system,
a local moment and a conduction electron form a
tightly bound
singlet,
\begin{eqnarray}
|s\rangle_i = (1/\sqrt{2})(|\!\uparrow\rangle_f|\!\downarrow\rangle_c
-
|\!\downarrow\rangle_f|\!\uparrow\rangle_c ),
\label{QS:tight-singlet}
\end{eqnarray}
with a large binding energy of order $J_K$.
Each of the remaining $(1-x)N_{\rm site}$ sites hosts a lone
local
moment
which, when projected 
onto
the low energy subspace, is written as
\begin{eqnarray}
|{\rm lone~
local~
moment}~\sigma\rangle_i = (-\sqrt{2}\,\sigma) c_{i,\bar{\sigma}}|s\rangle_i .
\label{QS:lone-moment}
\end{eqnarray}
We can take 
$|s\rangle_i$ as the vacuum state
at site $i$,
in which case
a lone
local
moment behaves as a hole with infinite repulsion\footnote{There is only
one conduction electron in the singlet.}
but with a kinetic energy
of order $W$~\cite{QS:Hewson}.
In the paramagnetic
phase, we can invoke Luttinger's theorem to conclude that the Fermi
surface encloses $(1-x)$ holes or, equivalently,
$(1+x)$ electrons per unit cell. This is the heavy fermion state in which
local moments, through an entanglement with conduction electrons,
participate in the
electron
fluid~\cite{QS:Hewson}.
The Fermi surface is large in this sense.
By continuity, the above considerations apply to the paramagnetic
part of the phase diagram with more realistic parameters.
We label this phase as ${\rm P_L}$, with the subscript denoting
a large Fermi surface.

The continuity argument is supported by explicit microscopic calculations.
In the
regime $I \ll J_K \ll W$, various approaches, in particular the slave-boson
mean-field 
theory, give rise to the following picture.
Consider the conduction electron Green's function,
\begin{eqnarray}
G_c({\bf k},\omega) \equiv F.T.[-\langle T_{\tau} c_{{\bf k},\sigma}(\tau)
c_{{\bf k},\sigma}^{\dagger}(0)\rangle] ,
\label{QS:gc-definition}
\end{eqnarray}
where 
the Fourier transform 
$F.T.$ is taken with respect to $\tau$. This Green's function
is related to a self-energy, $\Sigma({\bf k},\omega)$, via the standard
Dyson equation,
\begin{eqnarray}
G_c({\bf k},\omega) = \frac{1}{\omega-\varepsilon_{\bf k} -
\Sigma({\bf k},\omega)} .
\label{QS:gc-Dyson-equation}
\end{eqnarray}
In the heavy Fermi liquid state,
$\Sigma({\bf k},\omega)$ is non-analytic and contains
a pole in the energy space,
\begin{eqnarray}
\Sigma({\bf k},\omega)
=\frac{(b^*)^2}{\omega-\varepsilon_f^*} .
\label{QS:sigma-pole}
\end{eqnarray}
Inserting 
Eq.~(\ref{QS:sigma-pole})
into
Eq.~(\ref{QS:gc-Dyson-equation}), we end up with two poles
in the conduction electron Green's function,
\begin{eqnarray}
G_c({\bf k},\omega) =
\frac{u_{\bf k}^2}{\omega-E_{1,{\bf k}}}
+
\frac{v_{\bf k}^2}{\omega-E_{2,{\bf k}}} ,
\label{QS:gc-two-poles}
\end{eqnarray}
where
\begin{eqnarray}
E_{1,{\bf k}} &=& (1/2)[\varepsilon_{\bf k}+\varepsilon_f^*
-\sqrt{(\varepsilon_{\bf k}-\varepsilon_f^*)^2+4(b^*)^2} ] ,
\nonumber \\
E_{2,{\bf k}} &=&
(1/2)[\varepsilon_{\bf k}+\varepsilon_f^*
+\sqrt{(\varepsilon_{\bf k}-\varepsilon_f^*)^2+4(b^*)^2} ] .
\label{QS:gc-two-poles2}
\end{eqnarray}
These two poles describe the dispersion of the two heavy-fermion bands.
These bands must accommodate $1+x$ electrons, so the new Fermi energy
must lie in a relatively flat portion of the dispersion, leading
to a small Fermi velocity and a large 
quasi-particle 
mass $m^*$.

It is important to note that we have used a
${\bf k}$-independent self-energy to describe a
large reconstruction of the 
quasi-particle 
dispersion
($\varepsilon_{\bf k}$ $\to$
$E_{1,{\bf k}}, E_{2,{\bf k}}$) and a corresponding large
reconstruction of the Fermi surface. In fact,
the self-energy of Eq.~(\ref{QS:sigma-pole})
contains only two parameters,
the strength, i.e.,
the residue
of the pole, $(b^*)^2$, and the location of the pole,
$\varepsilon_f^*$. 
Equation
~(\ref{QS:sigma-pole}) does not contain the
incoherent features beyond the well-defined pole.
Such incoherent components can be introduced,
through, e.g.,
dynamical 
mean-field 
theory 
~\cite{QS:GeorgesRMP},
and they will
add
damping terms 
to
Eq.~(\ref{QS:gc-two-poles}).  
But the fact remains that a ${\bf k}$-independent
self-energy is adequate to capture the Kondo effect
and the resulting heavy 
quasi-particles. 
We will return to
this feature in the discussion of the Kondo breakdown effect.

\section{Quantum Criticality in the Kondo Lattice}
\label{QS:sec:qcp}

\subsection{General Considerations}

The Kondo interaction drives the formation of Kondo singlets between
the local moments and conduction electrons. At high temperatures,
the system is in a fully incoherent regime with the local moments
weakly coupled to conduction electrons. Going below some scale
$T_0$, the initial screening of the local moments starts to
set in. Eventually, at temperatures
below some Fermi-liquid scale, $T_{\rm FL}$, the heavy 
quasi-particles
are fully developed.

When the AF RKKY interaction among the local moments
becomes larger than
the Kondo interaction, the system is expected to develop
an AF order. An AF QCP is then to be expected
when the control parameter, 
$\delta = T_K^0/I$, 
reaches some
critical value $\delta_c$. At 
$\delta < \delta_c$,
the AF order
will develop as 
the 
temperature is lowered through the AF-ordering
line, $T_N(\delta)$.

In addition, 
RKKY interactions will eventually lead
to the suppression of the Kondo singlets. Qualitatively,
RKKY interactions promote singlet formation among
the local moments, thereby reducing the tendency of singlet
formation between the local moments and conduction electrons.
This will define an energy ($E_{\mathrm{loc}}^*$) 
or temperature ($T_{\mathrm{loc}}^*$)
scale, describing the breakdown of the Kondo effect. On very general
grounds, the $T_{\mathrm{loc}}^*$ line is expected to be a crossover
at non-zero temperatures, but turns into
a sharp transition at zero temperature.
The notion of Kondo breakdown in quantum critical heavy fermions
was introduced
in the theory of local quantum
criticality~\cite{QS:Si-Nature} and 
a 
related approach
based on fractionalization~\cite{QS:Colemanetal}. It 
also
appeared in 
subsequent work~\cite{QS:senthil2004a,QS:Pepin}
using
a gauge-theory formulation.
The Kondo breakdown effect is 
alternatively referred to as 
a Mott localization of the 
$f$-electrons.

\begin{figure}[t]
\centerline{\includegraphics*[width=0.8\linewidth,height=0.31\linewidth]{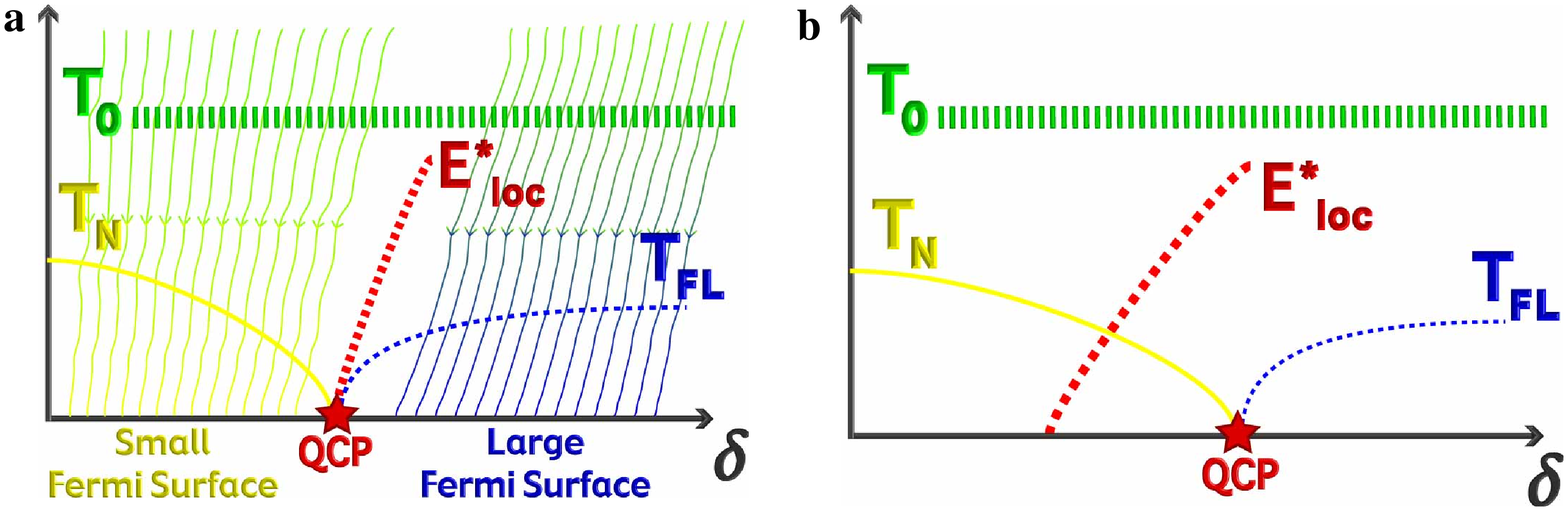}}
\vskip 0.3 cm
\caption{
Two types of 
QCPs
in Kondo lattice models~\cite{QS:Si-Nature}
(from Ref.~\cite{QS:Gegenwart.08}).
a) Locally-critical QPT.
A Kondo breakdown, signified by 
the vanishing of the energy scale $E_{\mathrm{loc}}^*$,
occurs at the continuous onset of AF order.
$T_{\mathrm{N}}$ is the AF transition temperature,
and $T_{\mathrm{FL}}$ is the temperature scale 
below which Fermi liquid behavior sets in.
$T_{\mathrm{0}}$ is a crossover temperature scale at which
Kondo screening initially sets in, and is also the upper bound
of the temperature range for quantum-critical scaling;
b) SDW QCP, where the Kondo breakdown does not occur until inside 
the region of AF order.
}\label{QS:lqcp}
\end{figure}

To study these issues theoretically, one key question is how
to capture not only the magnetic order and Kondo-screening, but also
the dynamical competition between the Kondo and RKKY interactions.
The microscopic approach
that is capable of doing this is the extended dynamical 
mean-field 
theory (EDMFT)~\cite{QS:Si.96,QS:SmithSi-edmft,QS:Chitra}.
The two solutions~\cite{QS:Si-Nature,QS:lcp-prb,QS:GrempelSi,QS:ZhuGrempelSi,QS:SunKotliar.03,QS:Glossop07,QS:Zhu07,QS:Glossop_etal09}
that have been derived through EDMFT are illustrated in Fig.~\ref{QS:lqcp},
and are summarized below.

Large-$N$ 
approaches based on slave-particle representations of the spin
operator are also commonly used to study Kondo-like systems.
One type of approach is based on a fermionic representation
of the 
spin~\cite{QS:senthil2004a}.
This representation
naturally incorporates the physics of singlet formation, so it
captures the Kondo singlets,
as well as the singlets among the local
moments, but it does not include magnetism in the
large-$N$ 
limit. One may allow a magnetic order in a static mean-field
theory for a 
finite-$N$
~\cite{QS:senthil2004a}. However,
the magnetic transition and breakdown of Kondo screening are
always separated in the phase diagram and the zero-temperature
magnetic transition is still of the SDW type. This, we believe,
is a manifestation of the static nature of the 
mean-field 
theory.

A Schwinger-boson-based 
large-$N$ 
formulation is another microscopic
approach that is being considered in this context~\cite{QS:Rech}.
This approach naturally incorporates
magnetism. While it is traditionally believed that bosonic
representations of spin in general have difficulty 
in capturing
the Kondo screening physics at its 
large-$N$ 
limit, there is
indication~\cite{QS:Rech} that the dynamical nature of the formulation
here allows an access to at least aspects of the Kondo effect.
It will be interesting to see what type of 
QPTs
this approach will lead to for the Kondo lattice problem.

Another approach~\cite{QS:Yamamoto07} to the Kondo lattice Hamiltonian
is based on a quantum non-linear sigma model (QNL$\sigma$M)
representation of the local moments. This approach is most readily
applied to the limit of $J_K\ll I \ll W$, the limit of large $\delta$.
To access the Kondo regime requires the incorporation
of 
Berry-phase 
terms in the representation, and such a study remains
to be carried out.

\subsection{Microscopic Approach Based on the Extended Dynamical Mean-field Theory}

The EDMFT method~\cite{QS:Si.96,QS:SmithSi-edmft,QS:Chitra}
incorporates 
inter-site 
collective fluctuations
into the 
dynamical mean-field theory
framework
\cite{QS:GeorgesRMP}.
The
systematic method is constructed within
a 
cavity, diagrammatic, or functional
formalism~\cite{QS:Si.96,QS:SmithSi-edmft,QS:Chitra}.
It is conserving, satisfying the various Ward identities.
Diagrammatically, EDMFT
incorporates an infinite series
associated with 
inter-site 
interactions,
in addition to
the local processes already taken into account in the
dynamical mean-field theory.

Within 
EDMFT, the dynamical spin susceptibility
and the conduction-electron Green's function
respectively have the forms
$\chi ({\bf q}, \omega) = [ M(\omega) + I_{{\bf q}}]^{-1} $,
and
$G ({\bf k}, \varepsilon) =
[\varepsilon + \mu - \varepsilon_{\bf k} - \Sigma (\varepsilon)]^{-1} $.
The correlation functions,
$\chi ({\bf q}, \omega)$ and
$G ({\bf k}, \varepsilon)$, are momentum-dependent.
At the same time, the irreducible quantities,
$ M(\omega)$ and
$\Sigma (\varepsilon)$,
are momentum-independent.
They are determined in terms of
a
Bose-Fermi Kondo model,
\begin{eqnarray}
{\cal H}_{\text{imp}} &=& J_K ~{\bf S} \cdot {\bf s}_c +
\sum_{p,\sigma} E_{p}~c_{p\sigma}^{\dagger}~ c_{p\sigma}
\nonumber \\
&& + \; g \sum_{p} {\bf S} \cdot \left( {\bf \Phi}_{p} +
{\bf \Phi}_{-p}^{\;\dagger} \right) +
\sum_{p}
w_{p}\,{\bf \Phi}_{p}^{\;\dagger} \cdot {\bf \Phi}_{p}\;.
\label{QS:H-imp}
\end{eqnarray}
The fermionic ($c_{p\sigma}$) and bosonic 
(${\bf \Phi}_{p}$)
baths are determined by self-consistency
conditions, which manifest the translational invariance,
$\chi_{{loc}} (\omega)
= \sum_{\bf q} \chi ( {\bf q},
\omega )$,
and $G_{{loc}} (\omega) = \sum_{\bf k} G( {\bf k}, \omega )$.
The 
$(0+1)$-dimensional 
quantum impurity problem,
Eq.~(\ref{QS:H-imp}), has the following Dyson equations:
$M(\omega)=\chi_{0}^{-1}(\omega) + 1/\chi_{\rm loc}(\omega)$
and $\Sigma(\omega)=G_0^{-1}(\omega) - 1/G_{\rm loc}(\omega)$, where
$\chi_{0}^{-1} (\omega) = -g^2 \sum_p 2 w_{p} /(\omega^2 -
w_{p}^2)$
and $G_0 (\omega) = \sum_p 1/(\omega - E_p)$ are the
Weiss fields.
The EDMFT formulation allows us to study different degrees of quantum
fluctuations as manifested in the spatial dimensionality of these
fluctuations. The case of two-dimensional
magnetic fluctuations is represented in terms of
the RKKY density of states that has a non-zero value at the lower edge,
e.g.,
\begin{eqnarray}
\rho_{I} (x) \equiv  \sum_{\bf q} \delta ( x  - I_{\bf q} )
= (1/{2 I}) \Theta(I - | x | ) \;,
\label{QS:rho_I_2D}
\end{eqnarray}
where $\Theta$ is the Heaviside step function.
Likewise, three-dimensional magnetic fluctuations are described
in terms of a $\rho_{I} (x)$ which vanishes at the lower edge
in a square-root fashion, 
e.g.,
\begin{eqnarray}
\rho_{I} (x) = (2/{\pi I^2}) \sqrt{I^2-x^2}\,
\Theta(I - | x | ) \;.
\label{QS:rho_I_3D}
\end{eqnarray}

The bosonic bath 
captures
the effect of the dynamical magnetic
correlations, primarily among the local moments, on the local Kondo
effect. As a magnetic QCP is approached, the spectrum of the
magnetic fluctuations softens, and so does that of the bosonic
bath. Consequently, its ability to suppress the Kondo effect
increases. This effect has been explicitly seen in a number
of specific studies
\cite{QS:Si-Nature,QS:lcp-prb,QS:GrempelSi,QS:ZhuGrempelSi,QS:SunKotliar.03,QS:Glossop07,QS:Zhu07,QS:Glossop_etal09}.
Moreover, the zero-temperature transition
is second-order whenever the same form of the
effective RKKY interaction appears in the formalism on both
sides of the transition~\cite{QS:SiGrempelZhu,QS:SunKotliar.05}.

\subsection{Spin-density-wave Quantum Critical Point}

The reduction of the Kondo-singlet amplitude
by the dynamical effects of 
RKKY interactions among the local
moments has been considered in some detail in a number of
studies based on EDMFT
\cite{QS:Si-Nature,QS:lcp-prb,QS:GrempelSi,QS:ZhuGrempelSi,QS:SunKotliar.03,QS:Glossop07,QS:Zhu07,QS:Glossop_etal09}.
Irrespective of the spatial dimensionality, this weakening of the
Kondo effect is seen through the reduction of the $E_{\mathrm{loc}}^*$ scale.

Two classes of solutions emerge depending on whether this
Kondo breakdown scale vanishes at the AF QCP.
In the case of Eq.~(\ref{QS:rho_I_3D}), $E_{\mathrm{loc}}^*$ has not yet been
completely suppressed to zero when the AF QCP,
$\delta_c$, is reached from the paramagnetic
side.\footnote{However, it can go to zero inside the AF region, 
as further discussed in Sec.~\ref{QS:sec:af_kl}.}
This is illustrated in Fig.~\ref{QS:lqcp}(b).
The quantum critical behavior, at energies below
$E_{\mathrm{loc}}^*$, falls within the
spin-density-wave
type
\cite{QS:Hertz,QS:Millis,QS:Moriya}.
The zero-temperature dynamical spin susceptibility has the
following form:
\begin{eqnarray}
\chi({\bf q}, \omega ) =
\frac{1}{f({\bf q}) - ia \omega}
\;.
\label{QS:chi-qw-sdw}
\end{eqnarray}
Here $f({\bf q})=I_{\bf q}-I_{\bf Q}$, and is generically $\propto
({\bf q}-{\bf Q})^2 $ as the wavevector ${\bf q}$ approaches
the AF ordering wavevector ${\bf Q}$.
The QCP is described by a Gaussian fixed point. At non-zero temperatures,
a dangerously irrelevant operator invalidates the $\omega/T$
scaling~\cite{QS:Millis,QS:Moriya}.

\subsection{Local Quantum Critical Point}

Another class of 
solutions 
corresponds to $E_{\mathrm{loc}}^*=0$ already at $\delta_c$,
as shown in Fig.~\ref{QS:lqcp}(a).
It arises in the case of Eq.~(\ref{QS:rho_I_2D}), where the quantum critical
magnetic fluctuations are strong enough to suppress the Kondo effect.
The solution to the local spin susceptibility has the form
\begin{eqnarray}
\chi({\bf q}, \omega ) =
\frac{1}{f({\bf q}) + A \,(-i\omega)^{\alpha} W(\omega/T)}\;.
\label{QS:chi-qw-T}
\end{eqnarray}
This expression was derived~\cite{QS:Si-Nature,QS:lcp-prb}
within 
EDMFT studies,
through the aid of an $\epsilon$-expansion approach
to the Bose-Fermi Kondo model.
At the AF QCP,
the Kondo effect itself is critically destroyed.
The calculation of the critical exponent $\alpha$ is
beyond the reach of the $\epsilon$-expansion. In the
Ising-anisotropic case,
numerical calculations have found 
$\alpha\sim 0.7$~\cite{QS:GrempelSi,QS:Glossop07,QS:Zhu07,QS:Glossop_etal09}.

The breakdown of the Kondo effect not only affects magnetic
dynamics, but also influences the single-electron excitations.
As the QCP is approached from the paramagnetic side, the 
quasi-particle
residue $z_L \propto (b^*)^2$,
where $b^*$ is the strength of the pole
of $\Sigma({\bf k},\omega)$ [{\it cf.} Eq.~(\ref{QS:sigma-pole})],
goes to zero. The large Fermi surface turns critical.

The breakdown of the large Fermi surface 
implies
that the Fermi
surface will be small on the 
antiferromagnetically ordered 
side.
To consider this property further, we turn to
the Kondo effect inside the AF phase.

\begin{figure}[t]
\centerline{\includegraphics*[width=0.5\linewidth,height=0.27\linewidth]{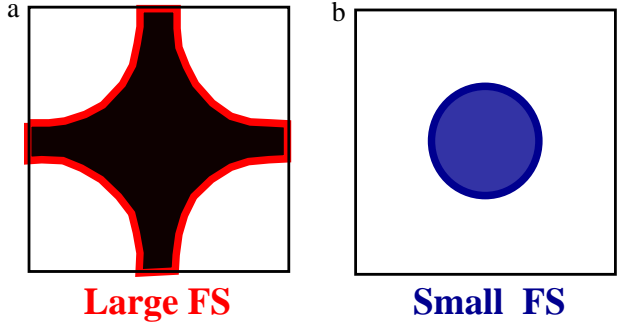}}%
\vskip 0.5 cm
\caption{
Fermi surface evolution across a local quantum critical point. 
As the system goes across the QCP 
from the paramagnetic ${\rm P_L}$ phase to the 
magnetically ordered ${\rm AF_S}$ phase,
a large Fermi surface 
[panel a)]
collapses 
and a small Fermi surface 
[panel b)]
emerges.
}    \label{QS:fermi_surface}
\end{figure}

\section{Antiferromagnetism and Fermi Surfaces in Kondo Lattices}
\label{QS:sec:af_kl}

To consider the Kondo effect in the 
antiferromagnetically ordered
phase,
we focus on the parameter regime of the Kondo lattice model,
Eq.~(\ref{QS:kondo-lattice-model}), in the
limit $J_K \ll I \ll W$.

In this 
parameter regime,
we can use as our reference point the $J_K=0$
limit
~\cite{QS:Yamamoto07}.
Then 
the local moments with AF exchange interactions
are decoupled from the conduction electrons.
We will focus on the case that the local-moment
system itself is in a collinear AF state.
Here, for low-energy physics, we can express the local-moment
 spin
density as
\begin{equation}
        {\bf S}_{\bf x}/s
=
\eta_{\bf x} {\bf n}_{\bf x} \sqrt{1-{\bf l}_{\bf x}^2}
+
{\bf l}_{\bf x} ,
\end{equation}
where
$s=\frac{1}{2}$ is the size of the local-moment
spin,
${\bf n}_{\bf x}$ and ${\bf l}_{\bf x}$
are the staggered and uniform components,
respectively,
and $\eta_x=\pm 1$ at even/odd sites;
consider, for definiteness,
a cubic or square lattice with N\'{e}el order.
The low-energy theory for the local-moment Hamiltonian,
the first term of Eq.~(\ref{QS:kondo-lattice-model}),
is the 
QNL$\sigma$M
\cite{QS:haldane1983,QS:Chakravarty}:
\begin{eqnarray}
{\cal S}_{\text{QNL}\sigma\text{M}}
=
(c/2g) \int d^dxd\tau\left[ \left(\nabla
{\bf n}
\right)^2 +
\left(
\partial{\bf n}/{ c ~\partial\tau}
\right)^2 \right]
\; .
\label{QS:qnlsm}
\end{eqnarray}
Here $c$ is the spin-wave velocity, and $g$
specifies the quantum
fluctuations.
There are gapless excitations in two regions of the wavevector
space: the staggered magnetization
(${\bf q}$ near ${\bf Q}$) specified by the ${\bf n}$
field and the uniform magnetization (${\bf q}$ near ${\bf 0}$)
described by ${\bf n} \times \partial {\bf n}/\partial \tau$.

When the Fermi surface of the
conduction electrons does not intersect the AF zone boundary,
only the uniform component of the local moments
can be coupled to the
spins of the conduction-electron states near the Fermi surface.
The effective Kondo coupling takes the following form,
\begin{eqnarray}
{\cal S}_K =
\lambda\int d^d{\bf x} d\tau
~
{\bf s}_c
\cdot
{\bf n} \times \partial {\bf n}/\partial \tau .
\label{QS:kondo-lambda}
\end{eqnarray}

A momentum-shell RG treatment requires a procedure
that mixes bosons, which scale along all directions in momentum
space,
and fermions, which scale along the
radial direction perpendicular to the Fermi surface~\cite{QS:shankar1994}.
Using the procedure specified in Ref.~\cite{QS:Yamamoto09},
we found $\lambda$ to be marginal at the leading
order
\cite{QS:Yamamoto07}, just like in the paramagnetic case.
The difference from the latter appears at the loop level:
$\lambda$ is exactly marginal to infinite loops~\cite{QS:Yamamoto07}.

The fact that $\lambda$ does not run towards infinity implies
a breakdown of the Kondo effect. This is supplemented
by a 
large-$N$ 
calculation~\cite{QS:Yamamoto07}, which showed
that the effective
Kondo coupling, Eq.~(\ref{QS:kondo-lambda}), leads to the following
self-energy for the conduction electrons:
\begin{eqnarray}
\Sigma({\bf k},\omega)
\propto \omega^d \; .
\label{QS:sigma-no-pole}
\end{eqnarray}
The absence of a pole in $\Sigma({\bf k},\omega)$, in contrast to
Eq.~(\ref{QS:sigma-pole}), implies the absence of any Kondo
resonance. Correspondingly, the Fermi surface is small.

\section{Towards a Global Phase Diagram}
\label{QS:sec:global}

\subsection{How to Melt a Kondo-destroyed Antiferromagnet}

Given the understanding that the AF state with a small Fermi surface
is a stable phase, it would be illuminating to approach the
quantum transition from this ordered state.

To do this, we must incorporate
the Berry-phase
term of the QNL$\sigma$M representation:
\begin{eqnarray}
{\cal S}_{\text{Berry}}
&=& i~s~\sum_{\bf x} \eta_{\bf x} A_{\bf x} ,
\nonumber\\
A_{\bf x} &=&
\int_0^{\beta}d\tau
\int_0^1 du
\left
[
{\bf n}\cdot
\left(\frac{\partial {\bf n}}{\partial u}
\times
\frac{\partial
{\bf n}}
{\partial \tau} \right) \right] .
\label{QS:berry-phase}
\end{eqnarray}
Here, 
$A_x$ is the area on the unit sphere spanned by
${\bf n}({\bf x},\tau)$ with $\tau \in (0,\beta)$.
The 
Berry-phase 
term can be neglected deep inside the AF phase.
For smooth configurations of ${\bf n}$ in the $({\bf x},\tau)$
space, the 
Berry-phase 
term vanishes. Topologically 
non-trivial
configurations of ${\bf n}$ in $({\bf x},\tau)$ yield a finite Berry
phase. 
However, 
they
cost a non-zero energy inside the AF phase and can
be neglected
for small $J_K$ and, correspondingly, small $\lambda$.
On the other hand,
as $J_K$ is increased these gapped configurations
come into play. Indeed, they are expected to be crucial for
capturing the Kondo effect.
Certainly, the Kondo singlet formation requires the knowledge of the size
of the microscopic spins, and the 
Berry-phase 
term is what encodes
the size of the spin in the QNL$\sigma$M representation.

\subsection{Global Phase Diagram}

\begin{figure}[t]%
\centerline{\includegraphics*[width=0.5\linewidth,height=0.36\linewidth]{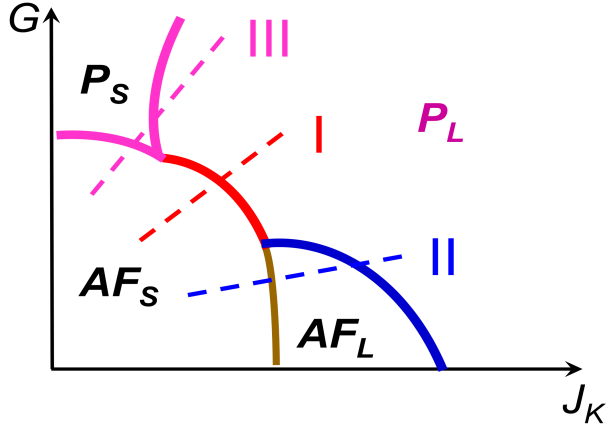}}%
\caption{%
The $T=0$ global phase diagram of the AF
Kondo lattice. $G$ describes the quantum fluctuations of the magnetic
Hamiltonian of the local moments, and $j_K$ is the normalized
Kondo coupling. 
${\rm P_L}$ and ${\rm P_S}$ respectively
describe paramagnetic phases with Fermi surfaces that are 
large and small, in the sense specified in the main text;
${\rm AF_L}$ and ${\rm AF_S}$ denote the corresponding
phases in the presence of an AF order.
(From Ref.~\cite{QS:Si_PSS.10},
and based on Ref.~\cite{QS:Si-physicab-06}.)
}
    \label{QS:global_pd}
\end{figure}

We can address these effects at a qualitative level, in terms of a
global phase diagram~\cite{QS:Si-physicab-06,QS:Si_PSS.10}.
We consider a two-dimensional
parameter space,
as shown in Fig.~\ref{QS:global_pd}.
The vertical axis describes the local-moment magnetism. It
is parametrized by $G$,
which characterizes the degree of quantum fluctuations
of the local-moment magnetism;
increasing $G$ reduces the N\'{e}el order.
This parameter can be a measure of magnetic frustration,
e.g.,
$G=I_{\rm nnn}/I_{\rm nn}$, the ratio of the
next-nearest-neighbor exchange interaction to the nearest-neighbor
one, or it can be the degree of spatial anisotropy.
The horizontal axis is $j_K\equiv J_K/W$, the Kondo
coupling normalized by the conduction-electron bandwidth.
We are considering a fixed value of $I/W$, which is typically
much less than $1$, and a fixed number of conduction electrons
per site, which is taken to be $0<x<1$
without
a loss of generality.

The ${\rm AF_S}$ phase describes the small-Fermi-surface
AF state, whose existence has been established
asymptotically exactly using the RG method as described in the
previous section.
The ${\rm P_L}$ phase is the standard heavy Fermi liquid
with heavy 
quasi-particles 
and a large Fermi surface~\cite{QS:Hewson}.
The ${\rm AF_L}$ phase corresponds to an AF state
in the presence of Kondo screening. It can either be considered
as resulting from the ${\rm AF_S}$ phase once the Kondo screening sets in,
or from the ${\rm P_L}$ phase via an SDW instability.
As alluded to in Ref.~\cite{QS:Si-physicab-06} and explicitly
discussed in Ref.~\cite{QS:Si_PSS.10},
a ${\rm P_S}$ phase should naturally arise,
describing a paramagnetic
phase with a Kondo breakdown
(and, hence, a small Fermi surface) which
either breaks or preserves translational invariance.
Related considerations are also being pursued in
Ref.~\cite{QS:Coleman_PSS.10}.

This global phase diagram contains three routes for a system to
go from the ${\rm AF_S}$ phase to the ${\rm P_L}$ phase.

\begin{itemize}

\item 
\textit{Trajectory I} is a direct transition between the two.
This ${\rm AF_S}-{\rm P_L}$ transition gives rise to a local 
QCP. 
A critical Kondo breakdown occurs at the
AF QCP. The Fermi surface undergoes
a sudden small-to-large jump of the Fermi surface
and the Kondo-breakdown scale $E_{\mathrm{loc}}^*$ vanishes
at the QCP~\cite{QS:Si-Nature,QS:Colemanetal}.
The 
quasi-particle 
residues associated with both the small
and large Fermi surfaces must vanish as the QCP is approached
from either
side.

\item 
\textit{Trajectory II} goes through the ${\rm AF_L}$ phase. The 
QCP 
at the ${\rm AF_L}-{\rm P_L}$
boundary falls in the 
spin-density-wave
type~\cite{QS:Hertz,QS:Millis,QS:Moriya}.
A Kondo breakdown transition can still take place at the
${\rm AF_L}-{\rm AF_S}$ boundary~\cite{QS:Si-physicab-06,QS:Yamamoto07}.

\item 
\textit{Trajectory III} goes through the ${\rm P_S}$ phase.
The ${\rm P_S}-{\rm P_L}$
transition could describe either a 
spin-liquid~\cite{QS:senthil2004a,QS:Anderson.08}
to heavy-Fermi-liquid
QCP,
or a 
spin-Peierls 
to heavy-Fermi-liquid 
QCP~\cite{QS:Pivovarov.04}.

\end{itemize}

\section{Experiments}
\label{QS:sec:expt-AF}

There 
has 
been considerable experimental work on quantum critical heavy
fermions. Here, we summarize a few points that are 
particularly 
pertinent
to the theoretical considerations discussed in this chapter. Readers are
referred to 
Chap.~18\cite{chapter:gegenwart}
for more details.

\subsection{Quantum Criticality}

The most direct evidence for the local quantum criticality
occurs in YbRh${\rm _2}$Si${\rm _2}$ and CeCu$_{\rm 6-x}$Au$_{\rm x}$.
For YbRh${\rm _2}$Si${\rm _2}$, the Fermi-liquid behavior is observed
both inside the AF-ordered phase and the field-induced non-magnetic
phase~\cite{QS:Custers.03}. In addition, Hall-coefficient measurements
\cite{QS:paschen2004,QS:friedemann_hall} have provided fairly direct evidence
for the breakdown of the Kondo effect precisely at the AF QCP.
The existence of the Kondo-breakdown scale, $T_{\mathrm{loc}}^*$, has
also been seen in both the Hall~\cite{QS:paschen2004,QS:friedemann_hall}
and thermodynamic~\cite{QS:gegenwart2007} experiments.

For CeCu$_{\rm 6-x}$Au$_{\rm x}$, the unusual magnetic
dynamics~\cite{QS:Schroder}
observed near the $x =x_c \approx 0.1$
by early
neutron scattering measurements
is understood in terms of such a critical Kondo breakdown
in the form of local quantum criticality.
A divergent effective mass expected in this picture is consistent
with the thermodynamic measurement in both the doping and
pressure-induced 
QCP in this system~\cite{QS:HvL-RMP}.
This picture necessarily implies a Fermi-surface jump across
the QCP, as well as a Kondo-breakdown energy scale
$E_{\mathrm{loc}}^*$ going to zero at the QCP, but such characteristics are
yet to be probed in CeCu$_{\rm 6-x}$Au$_{\rm x}$.

CeRhIn$_{\rm 5}$ is a member of the Ce-115 heavy 
fermions~\cite{QS:Hegger.00}.
It contains both antiferromagnetism and superconductivity
in its pressure-field phase diagram.
When a large-enough magnetic 
field 
is applied and superconductivity
is removed ($H>H_{c2}$), there is evidence for a single
QCP between antiferromagnetic 
and 
non-magnetic
phases~\cite{QS:park-nature06}.
At this QCP, the de Haas-van Alphen (dHvA) 
results~\cite{QS:shishido2005}
suggest a jump
in the Fermi surface and a divergence in the effective mass.
CeRhIn$_{\rm 5}$,
together with
$\beta$-YbAlB$_{\rm 4}$ \cite{QS:Nakatsuji.09},
illustrates the possibility
that local quantum criticality induces 
superconductivity.

One of the earliest systems in which anomalous magnetic dynamics
was observed is ${\rm UCu_{5-x}Pd_x}$~\cite{QS:Aronson.95}.
It is tempting to speculate~\cite{QS:Si-physicab-06} that a Kondo-destroying
spin-glass QCP underlies this observation.

There are also several heavy fermion systems in which spin-density-wave
type QCPs have been implicated.
Examples are
Ce(Ru${\rm _{1-x}}$Rh$_{\rm x}$)$_{\rm 2}$Si${\rm _2}$~\cite{QS:Kadowaki.06}
and Ce${\rm _{1-x}}$La$_{\rm x}$Ru$_{\rm 2}$Si${\rm _2}$~\cite{QS:Knafo.09}.

\subsection{Global Phase Diagram}

A number of heavy fermion materials
might be classified according to our global
phase diagram, Fig.~\ref{QS:global_pd}.

Perhaps the most complete information exists in the pure and
doped YbRh$_{\rm 2}$Si$_{\rm 2}$
systems.
In 
pure YbRh$_{\rm 2}$Si$_{\rm 2}$,
strong evidence exists that
the field-induced transition goes along the
trajectory 
I (see below). A surprising recent development came
from experiments in the doped YbRh$_{\rm 2}$Si$_{\rm 2}$.
In 
Co-doped YbRh$_{\rm 2}$Si$_{\rm 2}$, the field-induced
transition seems to travel along trajectory 
II
~\cite{QS:Friedemann09}.
In the Ir-doped~\cite{QS:Friedemann09} and 
Ge-doped~\cite{QS:Custers.09}
YbRh$_{\rm 2}$Si$_{\rm 2}$,
on the other hand, the field-induced transition appears
to go along trajectory 
III.

In CeCu$_{\rm 6-x}$Au$_{\rm x}$, both the pressure- and
doping-induced QCPs show the characteristics
of local quantum criticality, accessed through trajectory 
I.
However, 
the field-induced QCP~\cite{QS:Stockert_CeCuAu_field} 
has the properties of an
SDW QCP. We interpret the field-tuning as taking the trajectory
II. 
It will be interesting to explore
whether an ${\rm AF_S}$-${\rm AF_L}$
boundary can be located as a function of magnetic field.

CeIn$_3$ is one of the earliest heavy fermion metals in which
an AF QCP was implicated~\cite{QS:Mathur98}. This system is cubic,
and we would expect it to lie in the small $G$ region of the
global phase diagram. Indeed, there is indication
that this cubic material displays 
an ${\rm AF_S}$-${\rm AF_L}$ Lifshitz
transition as a function of magnetic field~\cite{QS:Sebastian.09}.

It is to be expected that magnetic frustration will help reach
the ${\rm P_S}$ phase.
The heavy fermion system
${\rm YbAgGe}$ has a hexagonal lattice,
and,
indeed, 
there is some indication
that the ${\rm P_S}$ phase exists in this
system~\cite{QS:Canfield}; 
however,
lower-temperature measurements over an extended
field range will be needed to help establish the
detailed phase diagram.

\section{Summary and Outlook}
\label{QS:sec:sum}

We close this chapter with some general observations and a few
remarks on open issues.

\subsection{Kondo Lattice}

Kondo lattice systems provide a concrete 
context in which to 
study quantum magnetism in
a metallic setting. Even though the ground states at generic band fillings are
metallic, local moments are well-defined degrees of freedom.

We have emphasized two complementary views of a Kondo lattice system.
On the one hand, we can consider it as a lattice of local moments.
This view is advantageous for understanding the Kondo-screening effect in the
Kondo lattice system. It allows us to build on the insights
that have been gained in the extensive studies of the single-impurity
Kondo 
problem.

On the other hand, we can also represent a Kondo lattice 
in terms of
a spin-$1/2$ Heisenberg model of the local moments, which are magnetically
coupled to a conduction electron band. This view allows us to
take advantage of the 
understanding of 
the inherent quantum
fluctuations of an underlying insulating quantum magnet.
The coupling to conduction electrons introduces an additional
source of quantum fluctuations.

\subsection{Quantum Criticality}

In the transverse-field Ising model we considered at the beginning, the
quantum-disordered state 
at $T=0$ and $\delta > \delta_c$
has spins polarized along the transverse
direction. This polarization is not spontaneously generated, but is
instead 
induced
by the externally applied transverse field.

In a Kondo lattice system, the quantum-disordered state has a
different character. Here, the quantum coherence is established through
the Kondo effect. Kondo-singlet formation, while not breaking
any symmetry of the Hamiltonian, shares an important characteristics
of the usual symmetry breaking: it is spontaneously generated.
This makes it natural for a critical destruction of such a
Kondo singlet to create its own critical singularity. The interplay
of this type of singularity with that associated with a continuous
onset of 
AF
order is at the heart of the local quantum
critical and related theoretical approaches to the Kondo breakdown
effect.

\subsection{Global Phase Diagram}

Theoretical considerations of the global phase diagram are only at the
beginning stage. There is much room for concrete studies. As mentioned
earlier, one type of quantum fluctuations in a Kondo lattice system
is that associated with the local-moment component alone. These
fluctuations can be tuned in various ways, such as varying the
degree of magnetic frustration, or tuning the spatial
dimensionality~\cite{QS:Si-Nature,QS:Matsuda.10}. 
Another type of quantum fluctuations
is induced by the coupling of these local moments to the conduction
electrons. Whether these two types of quantum fluctuations act in a
similar fashion, or lead to different types of ground states, provides
a way of thinking about the global phase diagram of the Kondo lattice
system. It will be instructive to start from the 
antiferromagnetically ordered 
state, and reach the various types of quantum disordered states.

\subsection{Superconductivity}

Unconventional superconductivity is prevalent in the heavy fermion
systems. As we mentioned in the beginning, historical studies of such
unconventional superconductivity are based on Cooper pairing mediated by
antiferromagnetic paramagnons.
It will be important
to see how Kondo destruction physics influences superconductivity.
In local quantum criticality,
for instance, there is
a strong fluctuation of the Fermi surface, between 
large
and
small. The excitations underlying such fluctuations and the
associated non-Fermi liquid behavior may very well be a key ingredient
for the superconducting pairing.

\vspace{0.5cm} 

\noindent{\bf Acknowledgments} 
-- 
I am grateful to my collaborators 
for their insights and discussions;
more recent work has been done in collaboration with
P. Goswami,
K. Ingersent,
S. Kirchner,
J. Pixley,
J. Wu, S. Yamamoto, J.-X. Zhu, and L. Zhu.
I would like to thank F. Steglich and P. Gegenwart
for discussions during the preparation of this chapter.
The work has been in part supported by
the NSF Grant No. DMR-0706625 and the Robert A. Welch
Foundation Grant No. C-1411.

\end{document}